\documentclass[twocolumn,showpacs,aps,prd]{revtex4-1}

\usepackage{graphicx}
\usepackage{dcolumn}
\usepackage{amsmath}
\usepackage{epsfig}
\usepackage{subfigure}

\input babarsym

\newcommand{\BABARPubYear}    {10}
\newcommand{\BABARPubNumber}  {10-030, BAD \#2363, Version 21}

\newcommand{\SLACPubNumber} {14360}

\def\figurebox#1#2#3{%
    \def\arg{#3}%
    \ifx\arg\empty
    {\hfill\vbox{\hsize#2\hrule\hbox to #2{\vrule\hfill\vbox to #1{\hsize#2\vfill}\vrule}\hrule}\hfill}%
    \else
    {\hfill\epsfbox{#3}\hfill}%
    \fi}

\begin{document}

\preprint{\babar-PUB-\BABARPubYear/\BABARPubNumber} 
\preprint{SLAC-PUB-\SLACPubNumber} 


\title{
{\large \bf
Searches for the baryon- and lepton-number violating decays 
$B^0\rightarrow\Lambda_c^+\ell^-$, $B^-\rightarrow\Lambda\ell^-$, 
and $B^-\rightarrow\bar{\Lambda}\ell^-$}
}

%
\author{P.~del~Amo~Sanchez}
\author{J.~P.~Lees}
\author{V.~Poireau}
\author{E.~Prencipe}
\author{V.~Tisserand}
\affiliation{Laboratoire d'Annecy-le-Vieux de Physique des Particules (LAPP), Universit\'e de Savoie, CNRS/IN2P3,  F-74941 Annecy-Le-Vieux, France}
\author{J.~Garra~Tico}
\author{E.~Grauges}
\affiliation{Universitat de Barcelona, Facultat de Fisica, Departament ECM, E-08028 Barcelona, Spain }
\author{M.~Martinelli$^{ab}$}
\author{D.~A.~Milanes}
\author{A.~Palano$^{ab}$ }
\author{M.~Pappagallo$^{ab}$ }
\affiliation{INFN Sezione di Bari$^{a}$; Dipartimento di Fisica, Universit\`a di Bari$^{b}$, I-70126 Bari, Italy }
% ...
\collaboration{The \babar\ Collaboration}
\noaffiliation


\date{\today}
\begin{abstract}
Searches for $B$ mesons decaying to final states containing
a baryon and a lepton are performed, where the baryon is
either $\Lambda_c$ or $\Lambda$ and the lepton is
a muon or an electron. These decays violate both baryon and lepton number
and would be a signature of physics beyond the standard model. 
No significant signal is observed in any of the decay modes,
and upper limits in the range $(3.2-520) \times 10^{-8}$
are set on the branching fractions at the 90\% confidence level. 
\end{abstract}
\maketitle
\section{Introduction} \label{sec:introduction}
Observations show that the universe 
contains much more matter than antimatter~\cite{Coppi:2004za, Steigman:1976ev}. 
This suggests that there are processes that violate $CP$-symmetry
and baryon-number conservation~\cite{Sakharov:1967dj}. 
However, experimentally observed $CP$ violation, 
combined with the baryon-number violating
processes that are allowed by the standard model~\cite{Kuzmin:1985mm},
cannot explain the observed matter-antimatter asymmetry.

Baryon-number violation is a prediction of many unification theories~\cite{Georgi:1974sy,Fritzsch:1974nn},
but the proton decay rates predicted by many of these models have not been observed.
Stringent limits have been placed on the lifetime of the proton~\cite{Nakamura:2010zzi}.
The non-observation of proton decay has been used to constrain
baryon- and lepton-number violating decays involving 
higher-generation quarks and leptons~\cite{Hou:2005iu};
in that study, the upper limit on the branching fraction for 
$B^0\rightarrow\Lambda_c^+\ell^-$ is calculated to be $4\times 10^{-29}$,
where $\ell$ is a lepton.
No upper limits are calculated for $B^-$ decays to $\Lambda\ell^-$
or $\bar{\Lambda} \ell^-$.

We report the results of searches for the decays
$B^0 \rightarrow \Lambda_c^+ \ell^-$, $B^- \rightarrow \Lambda \ell^-$, and
$B^- \rightarrow \bar{\Lambda} \ell^-$, where the lepton is
a muon or an electron~\cite{chg_conj}.
Neither lepton number nor baryon number are conserved in these decays.
This is the first measurement of the branching fractions for these decays. 
An observation of any of these decay processes would be a sign of new physics. 


\section{\boldmath The \babar\ detector and dataset\label{sec:babar}}
The data used in this analysis were recorded with the \babar\ detector
at the \pep2\ asymmetric energy \epem\ storage ring. 
The data sample consists of 429.0 \invfb recorded at the \FourS resonance 
($\sqrt{s}$=10.58 \gevcc, where $\sqrt{s}$ is the center-of-mass (CM) energy of the \epem system).
The sample contains $(471\pm 3)\times 10^{6}$  \BB pairs.

The \babar\ detector is described in detail elsewhere~\cite{Aubert:2001tu}.
Charged particle momenta are measured in a tracking system consisting of a 
five-layer, double-sided silicon vertex tracker (SVT) and a 40-layer central
drift chamber (DCH), immersed in a 1.5 T axial magnetic field. Photon and
electron energies are  measured in a CsI(Tl) electromagnetic calorimeter (EMC).
The magnetic flux return for the solenoid (IFR), instrumented with resistive 
plate chambers or limited streamer tubes, provides muon identification.
Charged particle identification (PID) is also provided by an internally reflecting 
ring-imaging Cherenkov detector (DIRC) and the energy loss \dedx measured by the SVT
and DCH. 
Information from all these detectors is used in the particle identification.

Simulated Monte Carlo (MC) events are generated to study detector
effects. The detector response is modeled using the GEANT4 software 
package~\cite{Agostinelli:2002hh}. Large numbers of signal events are generated
for the six decay modes, assuming that the \B meson decays do not produce any
preferred polarization of the $\Lambda_c^+$ or $\Lambda$. This sample is 
referred to as the signal MC.
For background studies,
a large sample of \BB events is produced, with the \B mesons 
decaying according
to the measured branching fractions~\cite{Lange:2001uf}. The same procedure
is used to generate background samples 
for \epem annihilation to lighter quark-antiquark pairs ($u,d,s,c$).
These two samples are referred to as background MC.

\section{\boldmath Overview of analysis\label{sec:analysis}}

We identify $B$-meson candidates using two kinematic variables:
the difference between half the CM energy 
of the colliding beams and the measured CM 
energy of the \B candidate, \DeltaE; 
and the energy-substituted mass \mes of the \B candidate. 
In the calculation of \mes, the precise knowledge  
of the initial state energy is used to improve
the resolution on the calculated mass of the $B$ candidate:
\mes$= \sqrt{[(s/2 + \vec{p}_i \cdot \vec{p}_B)/E_i]^2 - |\vec{p}_B|^2}$, 
where $(E_i,\vec{p}_i)$ and $\vec{p}_B$ are respectively the four-momentum of the \epem system 
and the three-momentum of the $B$-meson candidate in the laboratory frame.
A region of phase space in these two variables, in which fits will be 
performed to extract the signal yield, is defined by the ranges
$-0.2<\DeltaE<0.2$\gev and $5.2<\mes<5.3$\gevcc. This is referred to
as the fitting region. The signal for true $B$ candidates for the studied decays
is centered around $\DeltaE=0$ ($\sigma \approx 16$\mev) and $\mes=5.279$\gevcc
($\sigma \approx 3$\mevcc), where $\sigma$ is the experimental resolution. 

In any search for a rare or new process, it is important to
minimize experimenter's bias. To do this, a blind analysis is performed. 
The kinematic region of phase space that would be populated by true
signal events is hidden during optimization of the candidate selection
criteria.
We exclude events within roughly $\pm 5\sigma$ of the signal peak in \mes and
$\pm 4 \sigma$ in \DeltaE.
The non-blinded region is referred to as the sideband region.
We define a region within $\pm 2.5\sigma$ of the signal peak in \mes
and \DeltaE as the signal region.

The signal yield is extracted with an unbinned
extended maximum likelihood fit.
The total probability distribution function (PDF) is 
a sum of PDFs for signal and background. Each of these 
PDFs is a product of PDFs describing the dependence on \mes and \DeltaE.
For the $\Lambda_c^+\ell^-$ modes,
additional discriminating power is gained from a three-dimensional
PDF, where the output from a neural net discriminator is used as the third
variable. This discriminator is defined in the next section.


\section{\boldmath Candidate selection and optimization\label{sec:selection}}
$\Lambda_c^+$ candidates are reconstructed through the decay mode 
$\Lambda_c^+\rightarrow pK^-\pi^+$, which has a 
branching fraction of $(5.0\pm 1.3)\times 10^{-2}$~\cite{Nakamura:2010zzi}. 
Other studies of \B decays to $\Lambda_c^+$~\cite{Majewski:2008if} 
show that including additional $\Lambda_c^+$ decay modes 
would add little sensitivity to this analysis.
$\Lambda$ candidates are reconstructed through the decay 
$\Lambda \rightarrow p\pi^-$, which has a branching 
fraction of $(63.9\pm 0.5)\times 10^{-2}$~\cite{Nakamura:2010zzi}.
The final state tracks for both the $\Lambda_c^+$ and $\Lambda$ decays are constrained 
to a common spatial vertex and their invariant mass is constrained to the 
$\Lambda_c^+$ or $\Lambda$ mass~\cite{Nakamura:2010zzi}.
This has the effect of improving the four-momentum
resolution for true $B\rightarrow \Lambda_{(c)}\ell$ candidates.

\B-meson candidates are formed by combining a $\Lambda_c^+$, $\Lambda$
or $\bar{\Lambda}$ candidate with a $\mu^-$ or $e^-$. 
The baryon and lepton candidates are constrained to originate
from a common point. The final state hadron ($p,K,\pi$) and lepton ($\mu,e$) 
candidates are all required to be consistent with the candidate particle
hypothesis according to PID algorithms that use \dedx, DIRC, EMC and IFR information.
The four-momenta of photons that are consistent with 
bremsstrahlung radiation from the electron candidate are
added to that of the electron.
As the $\Lambda$ has $c\tau=7.89$~\cm, the purity of the $\Lambda$-candidate sample
is improved by 
selecting candidates for which the reconstructed 
decay point of the $\Lambda$ candidate is at least $0.2$~\cm
from the reconstructed decay point of the \B candidate
in the plane perpendicular to the \epem beams.

A non-negligible background for the $\Lambda\ell$ channel is due
to incorrect identification of electrons and positrons in 
$e^+e^-\rightarrow e^+e^-\gamma$ events in which
the photon converts into an additional $e^+e^-$ pair and an electron
and positron in the final state are misidentified as a $\pi^-$ and proton coming from
a $\Lambda$ decay. 
This background is almost entirely eliminated by requiring that there are more than 
four charged tracks in the event. We apply this selection criterion
to all channels.

Candidate selection optimization is performed balancing two goals:
setting the lowest upper limit while remaining sensitive to a signal. 
We use the Punzi figure of merit (FOM)~\cite{Punzi:2003bu},
$\epsilon/(a/2 + \sqrt{N_{\rm bkg}})$, 
where $\epsilon$ is the signal efficiency,
$N_{\rm bkg}$ is the expected number of background events,
and $a$ is the number of standard deviations of 
significance at which the analysts 
would claim a discovery. For this analysis, $a=5$ is used. 
The signal efficiency and the expected number of background
events are obtained from the respective MC samples.

For optimization of the $p,K,\pi$ candidate PID selection, 
we calculate the Punzi FOM by estimating $\epsilon$ and $N_{\rm bkg}$
from the baryon candidate invariant mass distribution. 
The background is assumed to be linear in the baryon invariant mass
and a fit is made to extract the number of 
signal and background candidates.
After the PID selection is optimized, we select candidates
within $\pm 15$~\mevcc of the nominal $\Lambda_c^+$ mass
and $\pm 4$~\mevcc of the nominal $\Lambda$ mass~\cite{Nakamura:2010zzi}.

The lepton candidate selection is optimized based on the 
number of $B\rightarrow \Lambda_{(c)}\ell$ candidates in the signal region in MC samples.
These \B candidates contain 
correctly identified leptons from the signal decay, which define $\epsilon$
in the Punzi FOM, and two types of background, which determine $N_{\rm bkg}$: 
correctly identified leptons from standard model processes, 
and incorrectly identified leptons. 

A neural net is used to provide further discrimination 
between signal and background.
We use the {\sc TMVA} software package~\cite{Hocker:2007ht}
and its multilayer perceptron implementation of a neural net.
The neural net is trained using 
MC simulated samples for signal and for $e^+e^-\rightarrow q\bar{q}$ 
($q=u,d,s,c$).
The six discriminating variables, all defined in the CM frame of 
the \epem beams, used in the neural net algorithm are 
the angle between the \B meson momentum and the axis
defined by the colliding \epem system, 
the angle between the \B meson candidate sphericity~\cite{De Rujula:1978yh}
axis and the sphericity axis defined by the charged particles in the 
rest of the event (ROE), 
the angle between the \B meson candidate thrust~\cite{De Rujula:1978yh}
axis and the thrust axis defined by the charged particles in the ROE, 
the ratio of the $2^{\rm nd}$ to $0^{\rm th}$ 
Fox-Wolfram moment~\cite{Fox:1978vw} calculated
from the entire event using both charged and neutral particles,
$L$-moments of the ROE tracks~\cite{Lmoments}, and 
the magnitude of the thrust of the entire event.
For the $\Lambda\ell$ modes, the ratio of the Fox-Wolfram 
moments and the magnitude of the thrust of the entire event
show a slight correlation with \DeltaE and \mes 
in the background sample and are therefore not used.

For the $\Lambda_c^+\ell^-$ decay modes, we retain events with 
a value of the neural net output above a threshold such that 
about 90\% of the signal is retained and about 50\% of 
the background is rejected. 
The neural net output for the retained events
is used as a third discriminating variable in the PDF used in the fit. 

The $\Lambda\ell$ modes have significantly less background
than the $\Lambda_c^+\ell^-$ modes.
For the $\Lambda\ell$ modes, we retain events with a value of the neural net output above a threshold
(optimized using the Punzi FOM) and
perform a fit in \DeltaE and \mes only. 

After the optimized selection criteria are applied,
the remaining background for the $\Lambda_c^+\ell^-$ modes is composed of roughly 
equal amounts of \BB and $q\bar{q}$ ($q=u,d,s,c$) events, while the background 
for the $\Lambda\ell$ modes is almost entirely $q\bar{q}$.

\section{\boldmath Extraction of results\label{sec:extraction}}
As stated earlier, the signal yield is extracted with an unbinned
extended maximum likelihood fit.
For all decay modes, the signal \mes PDF is modeled 
as a Crystal Ball function~\cite{crystalballfunc}, which has
three free parameters. The signal \DeltaE PDF is the sum of two Crystal Ball functions
with the same mean.
For the $\Lambda_c^+\ell^-$ decay modes, the signal neural net output is 
modeled by a non-parametric PDF implemented in the {\tt RooFit}~\cite{Verkerke:2003ir} package
that models the distribution as a superposition of Gaussian
kernels~\cite{Cranmer:2000du}. The full signal PDF is a product of these PDFs. 
Signal MC samples for each decay mode are used to determine the parameter values for these functions, and
these values are fixed in the fit to the data. 

For all decay modes, the background \mes PDF is modeled 
as an ARGUS function~\cite{Albrecht:1986nr} and the background \DeltaE PDF 
is modeled as a linear function.
The unnormalized ARGUS function is defined as 
$\Psi(m)=m u^p e^{c u}$,
where $p=0.5$, $u=1-(m/m_0)^2$, $c$ is the curvature parameter, and 
$m_0$ is the kinematic cutoff above which the function is defined to be 0.
We determine $m_0=5.290$ \gevcc by a fit to the background MC events and fix this 
value in the fit to the data.
For the $\Lambda_c^+\ell^-$ decay modes, the background neural net output 
PDF is modeled as a Crystal Ball function. 

In the fit, the number of background events is a free parameter and the number
of signal events $S$ is the product of 
the branching fraction $\mathcal{B}$, which is treated 
as a free parameter, and a conversion factor $\mathcal{F}$:
$S=\mathcal{B}\mathcal{F}$, where
$\mathcal{F} = \epsilon\mathcal{B}_{\Lambda_{(c)}} N_{B}$,
$\mathcal{B}_{\Lambda_{(c)}}$ is the branching fraction for the 
$\Lambda_c^+$ or $\Lambda$, and $N_{B}$ is the number of 
either neutral or charged $B$ mesons in the dataset; 
$N_{B}=2N_{\Y4S}\mathcal{B}_{\BB}$, where 
$N_{\Y4S}$ is the number of $\Y4S$ in the dataset, 
$\mathcal{B}_{\BB}$ is the branching fraction for 
the \Y4S to decay to either a neutral or charged \BB pair,
and the factor of 2 accounts for the pair of $B$ mesons produced 
in each \Y4S decay.
There are no other free parameters for the signal PDF. 
The two-dimensional background PDF for the $\Lambda\ell$ modes has 
two free parameters (the \DeltaE slope and the \mes ARGUS shape parameter); 
the three-dimensional background PDF for the $\Lambda_c^+\ell^-$ modes
has three additional free parameters for the Crystal Ball function that models
the neural net output.

In order to incorporate systematic uncertainties on $\mathcal{F}$ (discussed below)
directly in the fit and propagate them to the total uncertainty on the branching fraction,
a Gaussian constraint is included as a term ($\mathcal{G}$) in
the $\ln$-likelihood function ($\ln\mathcal{L}$):
$\mathcal{G}=(\mathcal{F}-\mathcal{F}_{\rm fit})^2/2\delta_{\mathcal{F}}^2$,
where $\mathcal{F}$ is the value we calculate for the
conversion factor, $\mathcal{F}_{\rm fit}$ is a free parameter,
and $\delta_{\mathcal{F}}$ is the uncertainty on the conversion factor.
The Gaussian constraint is turned off in a subsequent fit to extract the statistical uncertainty 
only. This error is then subtracted in quadrature from the total 
error to determine the systematic error on the branching fraction
from these sources.

In order to test the stability and sensitivity of the fitting procedure 
as well as to search for possible bias in the fit, simulated signal events 
are embedded in a sample of background events generated from the background 
PDFs using MC techniques. We generate many independent samples with 
varying ratios of the number of embedded signal events to the number 
of background events in order to model different branching fractions.
These samples are fit and the extracted branching fractions are compared with 
the branching fractions used to determine the amount of embedded signal events. 
Biases of no more than 20\% of the statistical uncertainty on
the result of an individual fit are observed, 
depending on the decay mode and the number of signal events.
In addition, for $\mathcal{B}=0$ fits, we observe that 0.1\% to 1.5\% of fits 
(depending on decay mode) have no candidates in the signal region, and are 
thus unable to constrain the signal parameters. 
To avoid unmathematical negative PDF values~\cite{Porter:2003ui}, we refit these cases in the data
with a constraint that the PDF must be positive throughout the fitting region.

Systematic uncertainties are due to uncertainties on $\Lambda$ and $\Lambda_c^+$ 
branching fractions, the total number of
$B$ mesons produced during the experiment's lifetime, and the tracking and PID efficiencies,
which are determined from control samples in data.
We use the measured branching fractions and associated uncertainties for
$\Upsilon(4S)\rightarrow B^+B^-$ and $\Upsilon(4S)\rightarrow B^0\bar{B}^0$,
which are $(51.6\pm 0.6)\times 10^{-2}$ and $(48.4\pm 0.6)\times 10^{-2}$, respectively~\cite{Aubert:2005bq}.
For the $\Lambda_c^+\ell^-$ mode, the systematic uncertainty is dominated
by the 26\% uncertainty on the $\Lambda_c^+ \rightarrow p K^- \pi^+$ branching fraction;
the other uncertainties contribute about 3\%.
We do not assign any systematic uncertainty
due to the assumption of an unpolarized final state.
Systematic uncertainties from the fixed PDF parameters are considered to be negligible.
The total systematic uncertainties are estimated 
to be $26\%$ for $B^0\rightarrow\Lambda_c^+\ell^-$, 
$3.0\%$ for $B^-\rightarrow\Lambda\mu^-$ and $B^-\rightarrow\bar{\Lambda}\mu^-$,
and $2.5\%$ for $B^-\rightarrow\Lambda e^-$ and $B^-\rightarrow\bar{\Lambda} e^-$.

\begin{figure}[]
\begin{center}
\includegraphics[width=0.23\textwidth]{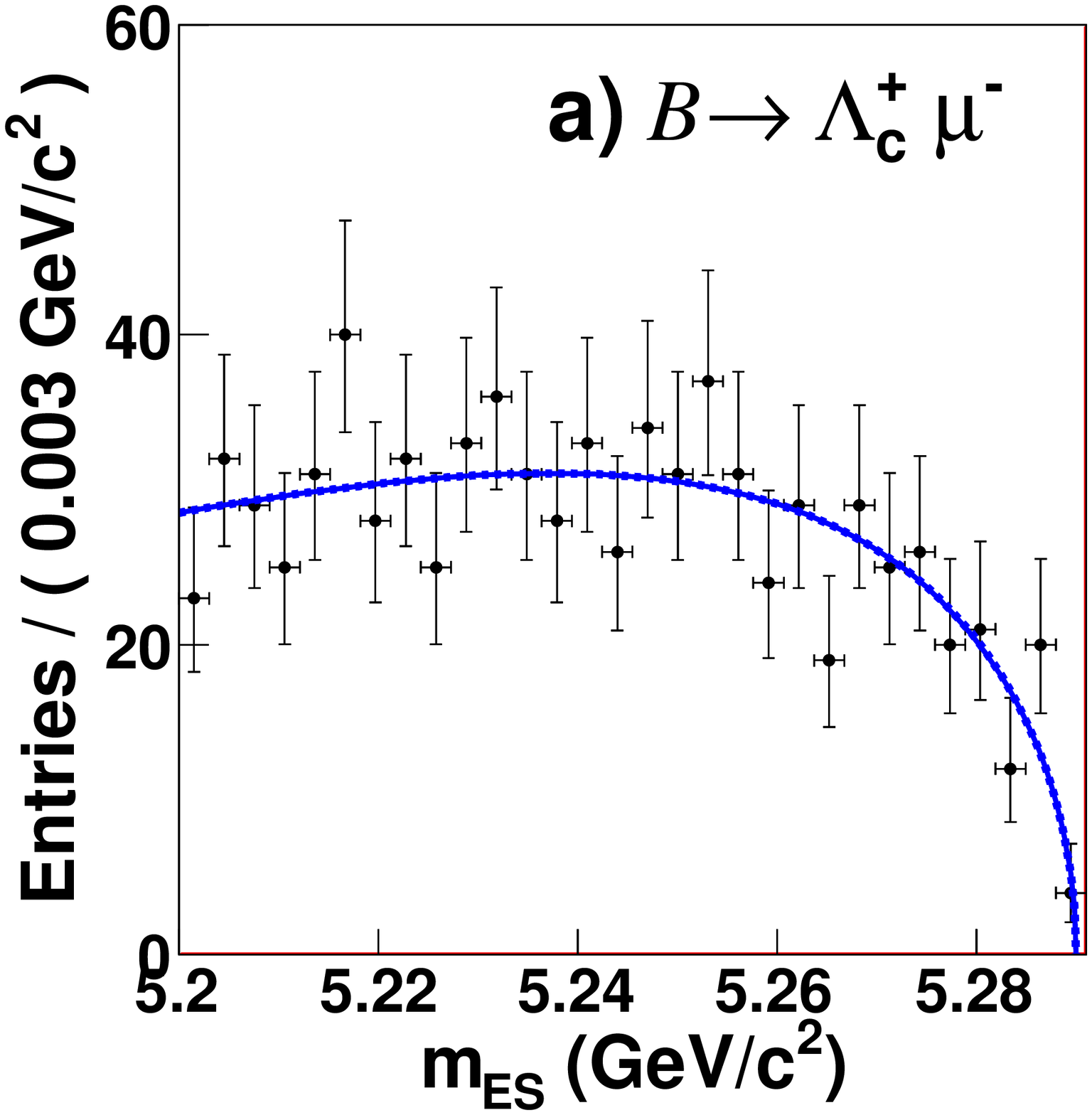}\hfill
\includegraphics[width=0.23\textwidth]{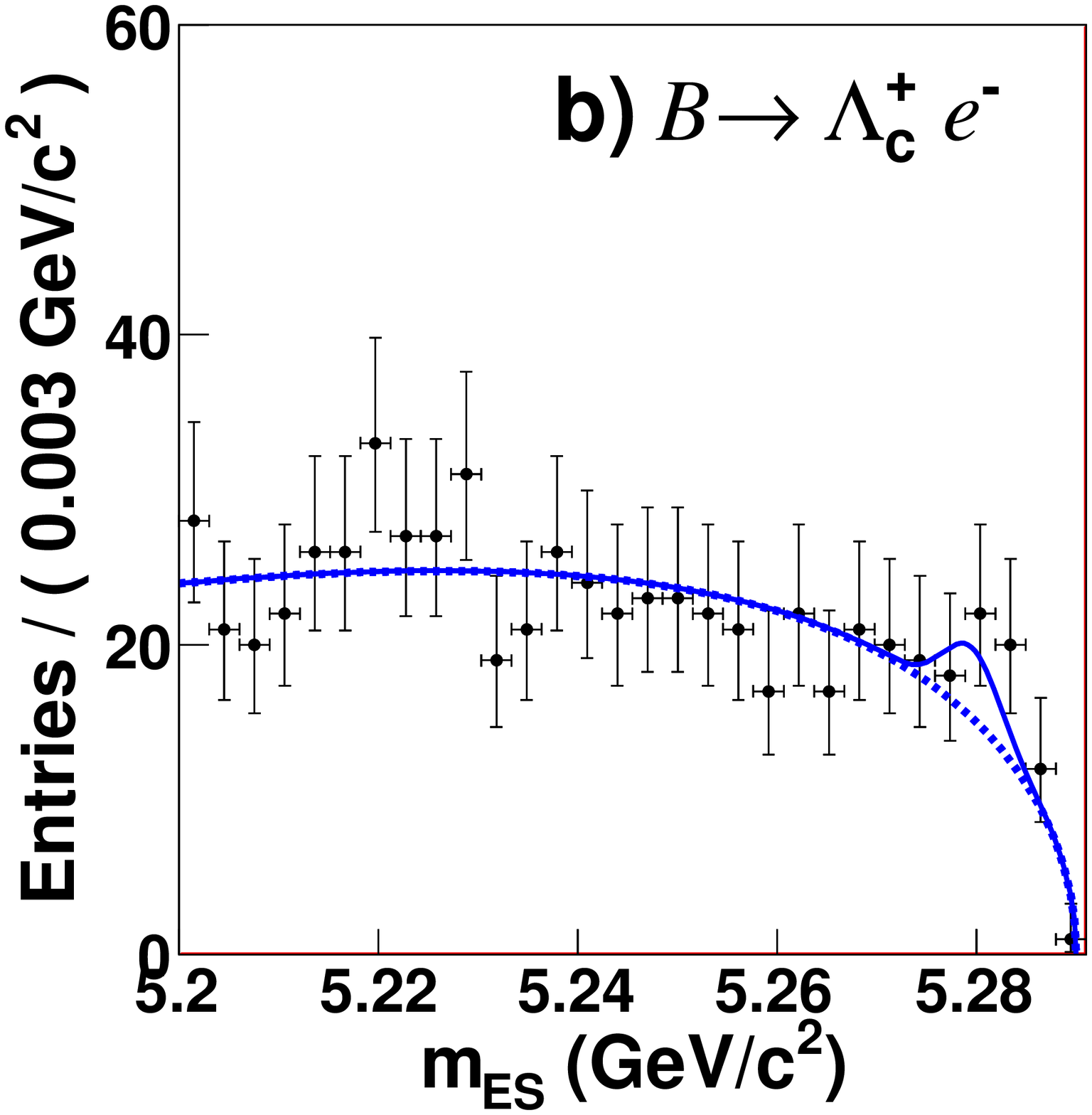}
\includegraphics[width=0.23\textwidth]{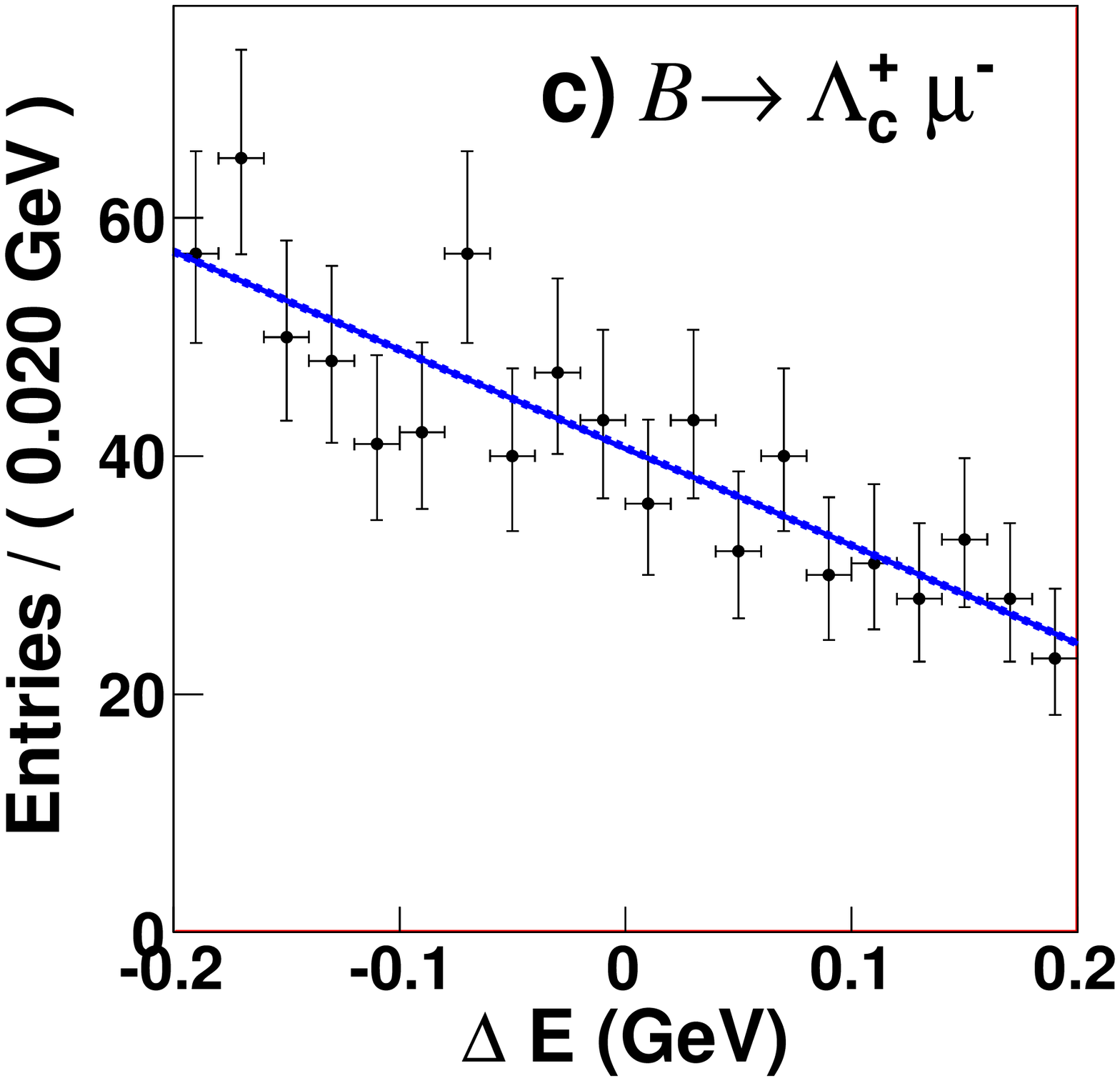}\hfill
\includegraphics[width=0.23\textwidth]{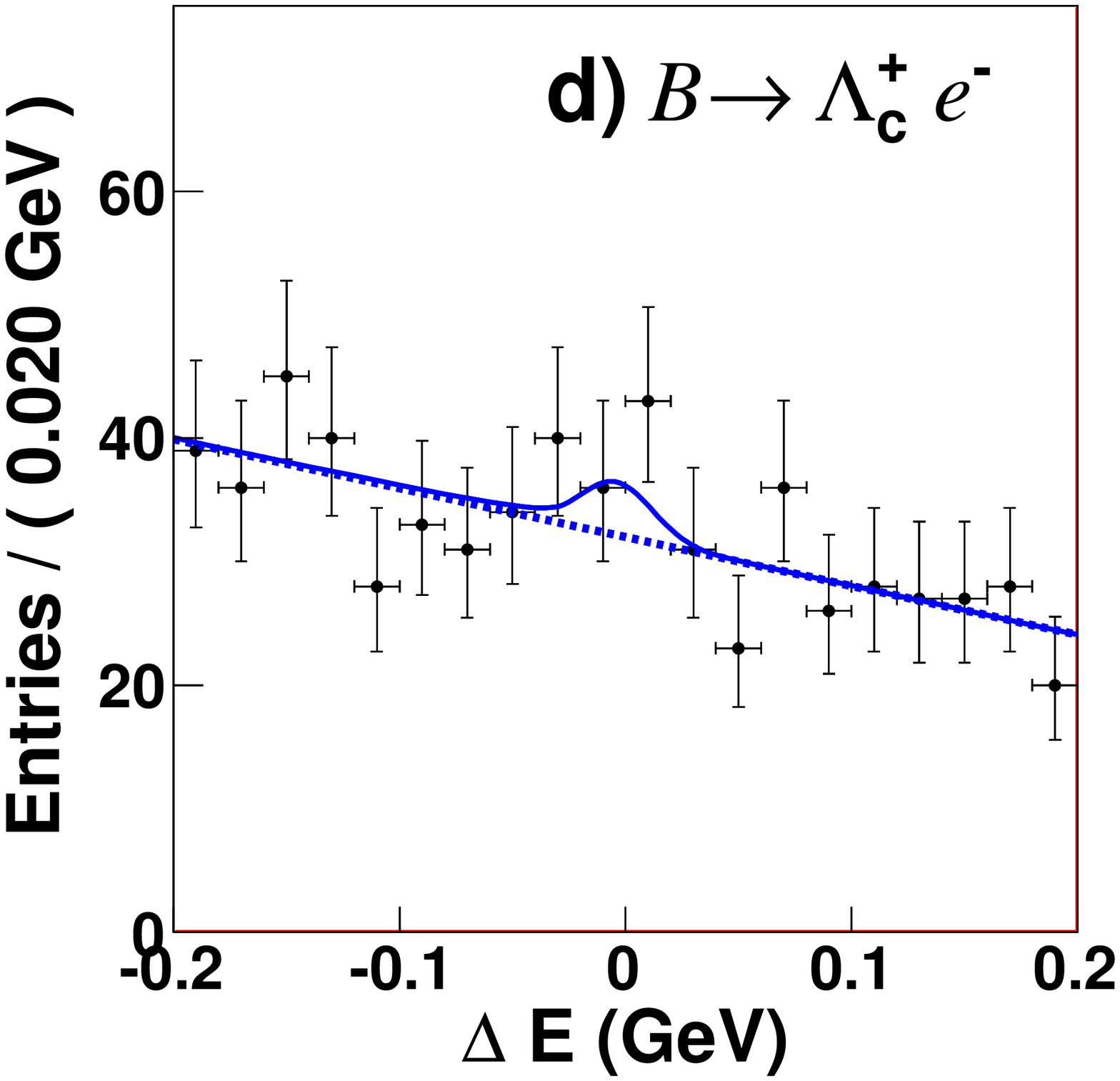}
\includegraphics[width=0.23\textwidth]{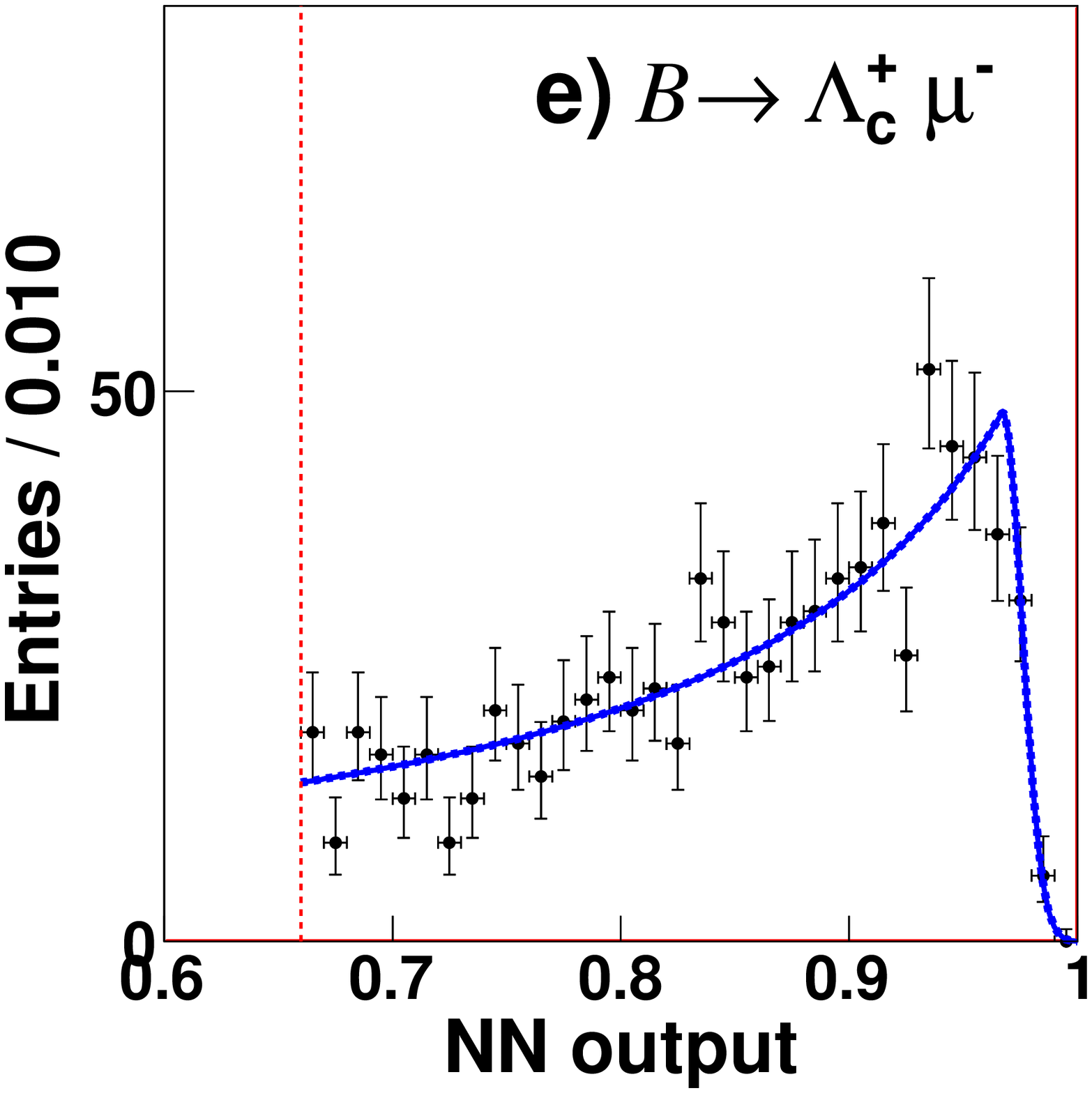}\hfill
\includegraphics[width=0.23\textwidth]{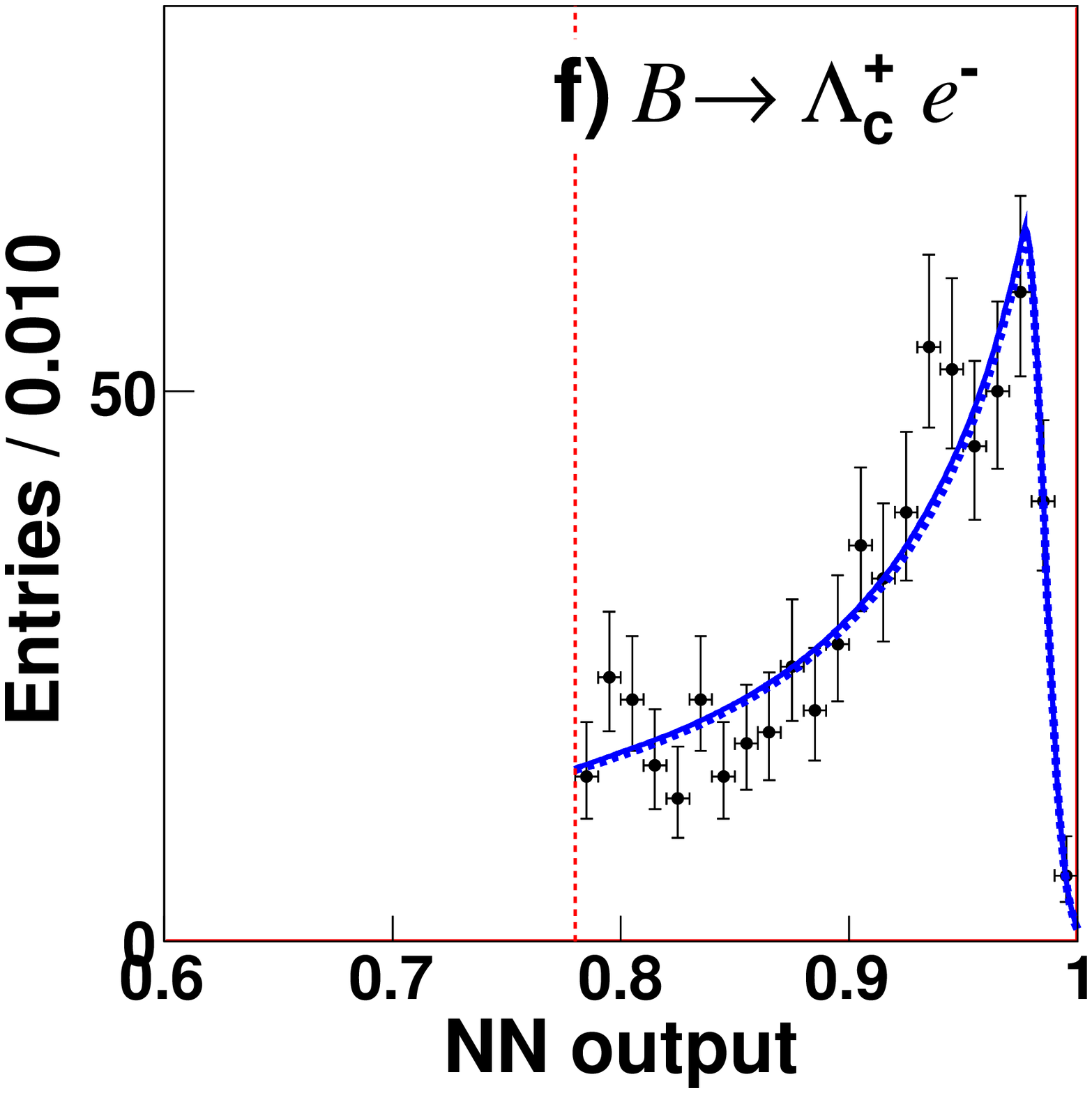}
\caption{Data for all events in the fitting region with overlaid fit results for $B^0\rightarrow\Lambda_c^+ \ell^-$ candidates. The left column
    is for the muon mode and the right column the electron mode.
    Distributions of (a,b) \mes, (c,d) \DeltaE, and (e,f)
    neural net (NN) output are shown. Dashed lines represent the background components of the fit and solid
    lines represent the sum of the signal and background components. The lack of a significant signal
    makes the solid and dashed lines indistinguishable in some plots. 
    In subfigure (e) and (f), the vertical dashed line 
    indicates the selection criteria on the neural net output.\label{fig:data_lc}}
\end{center}
\end{figure}

\begin{figure}[]
\begin{center}
\includegraphics[width=0.23\textwidth]{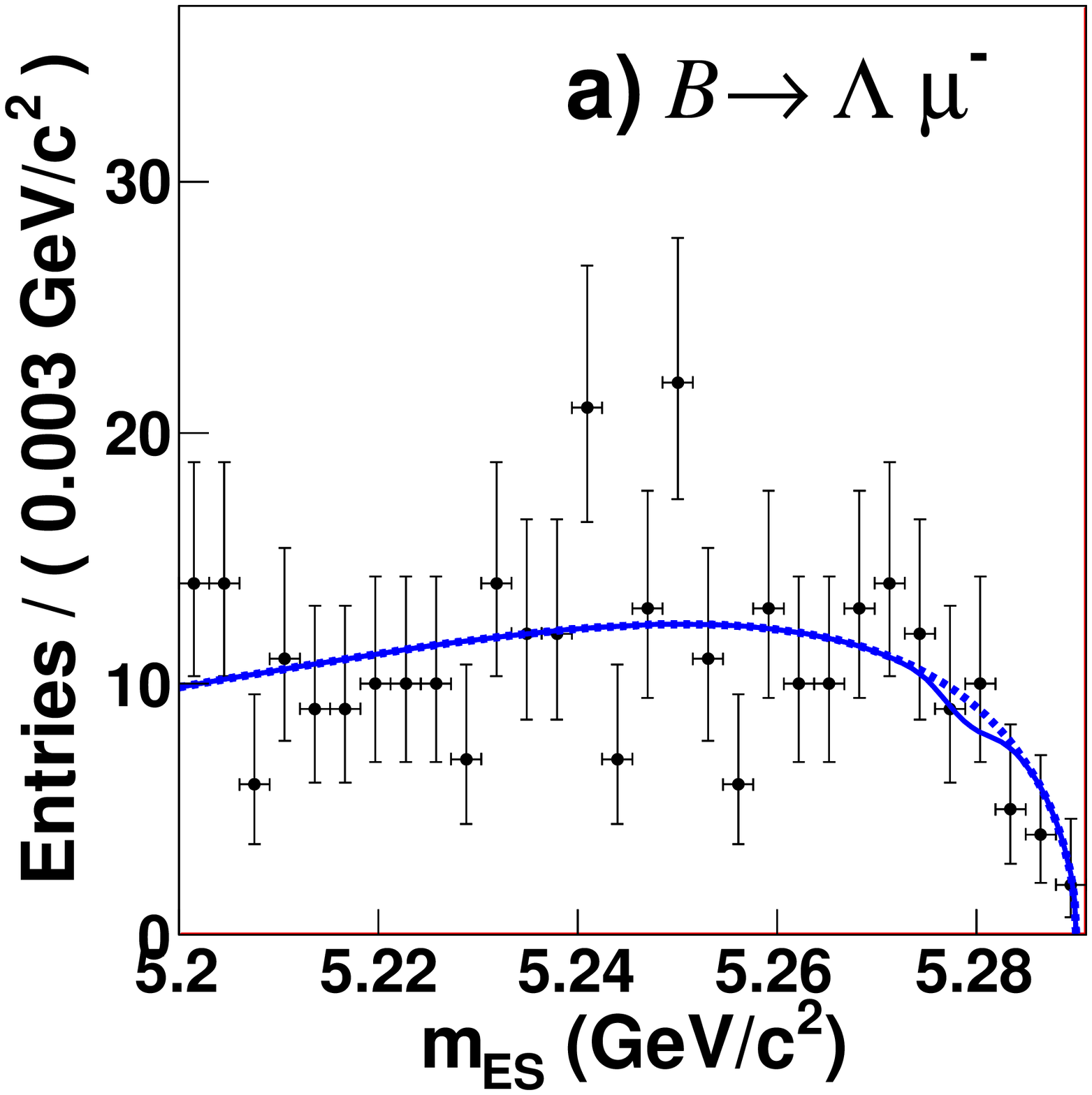}\hfill
\includegraphics[width=0.23\textwidth]{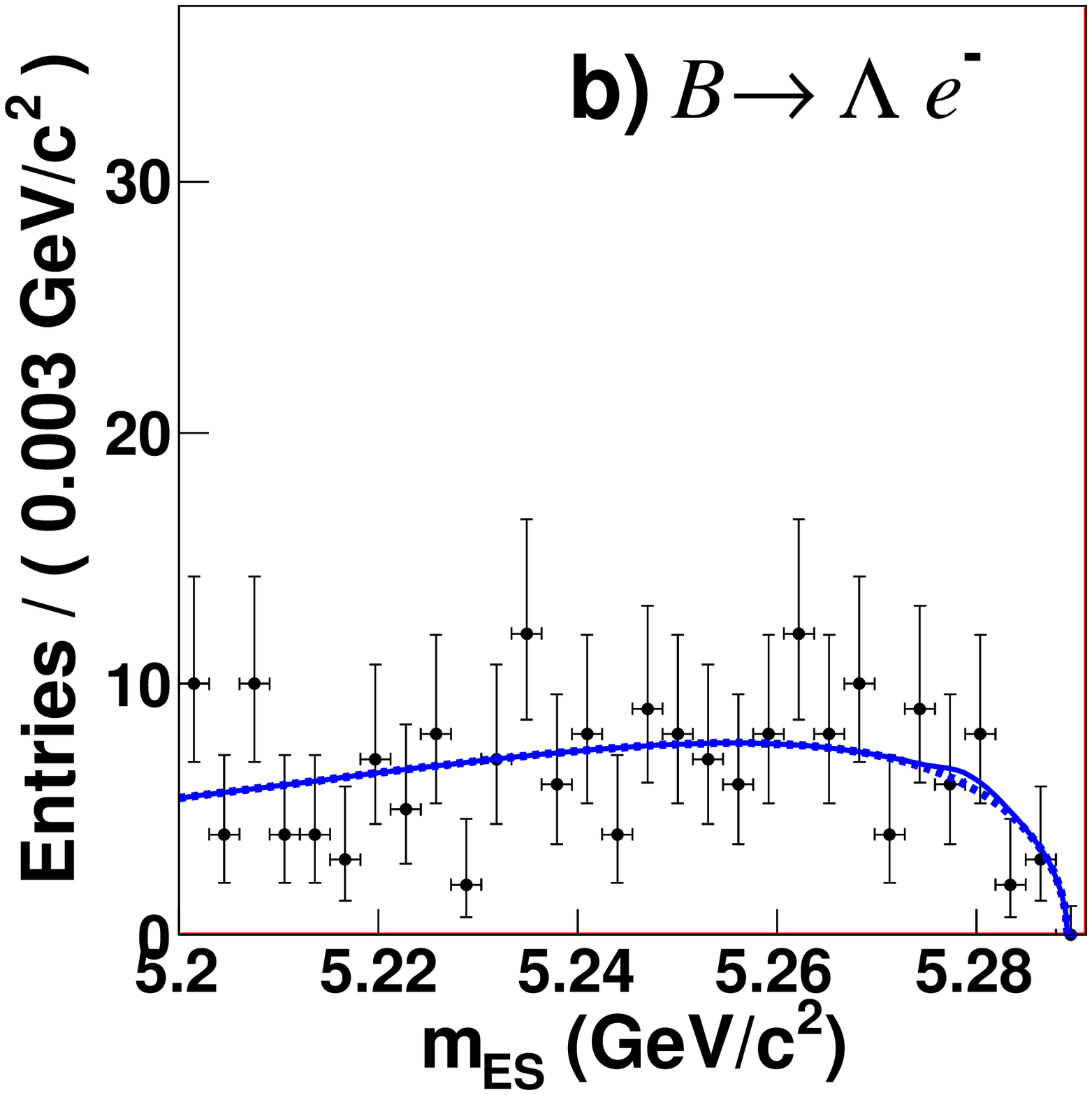}
\includegraphics[width=0.23\textwidth]{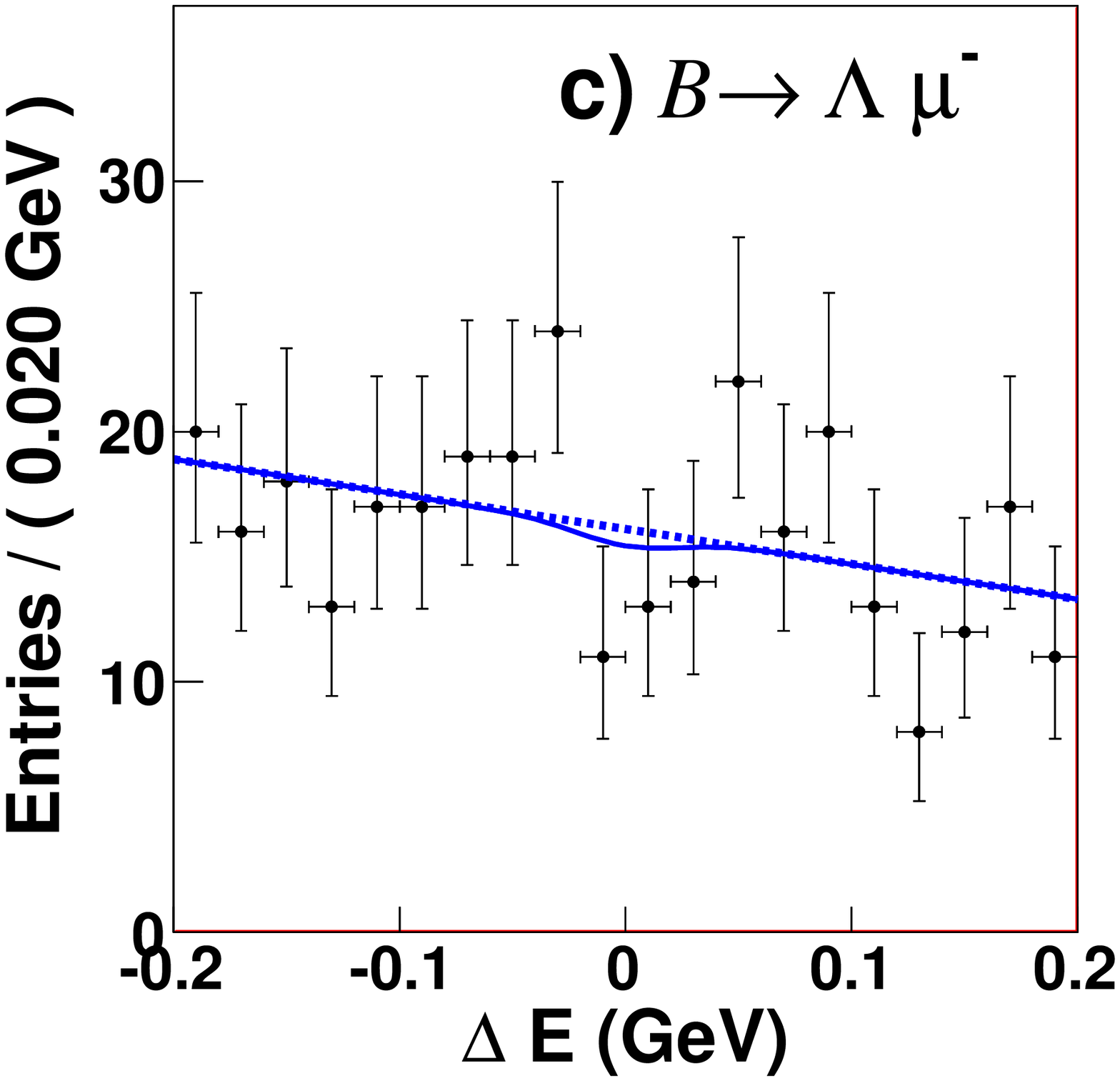}\hfill
\includegraphics[width=0.23\textwidth]{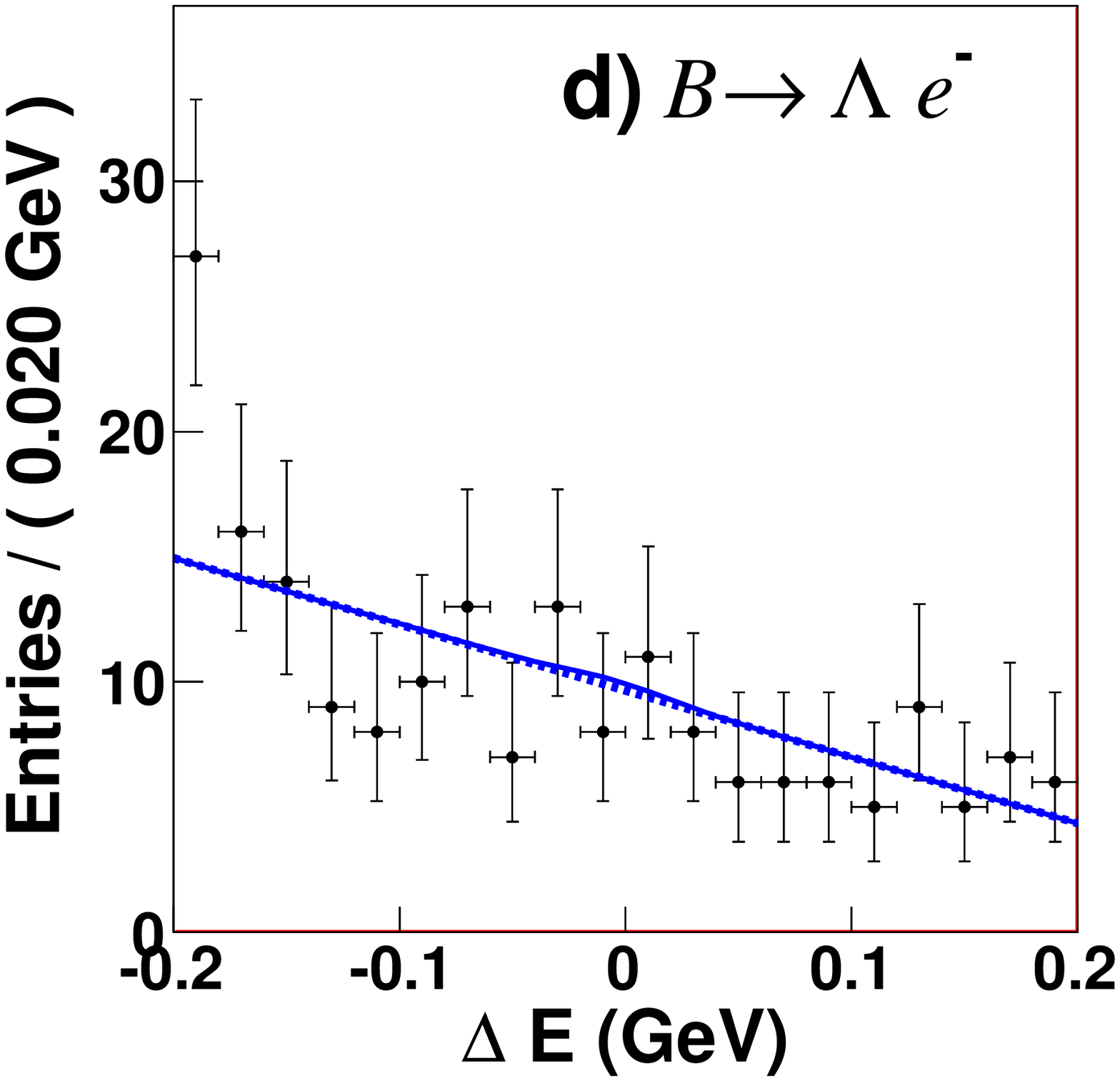}
\caption{Data for all events in the fitting region with overlaid fit results for $B^-\rightarrow\Lambda \ell^-$ candidates. The left column
    is for the muon mode and the right column the electron mode.
    Distributions of (a,b) \mes and (c,d) \DeltaE are shown.
    Dashed lines represent the background components of the fit and solid
    lines represent the sum of the signal and background components. The lack of a significant signal
    makes the solid and dashed lines indistinguishable in some plots.\label{fig:data_l0}}
\end{center}
\end{figure}

\begin{figure}[]
\begin{center}
\includegraphics[width=0.23\textwidth]{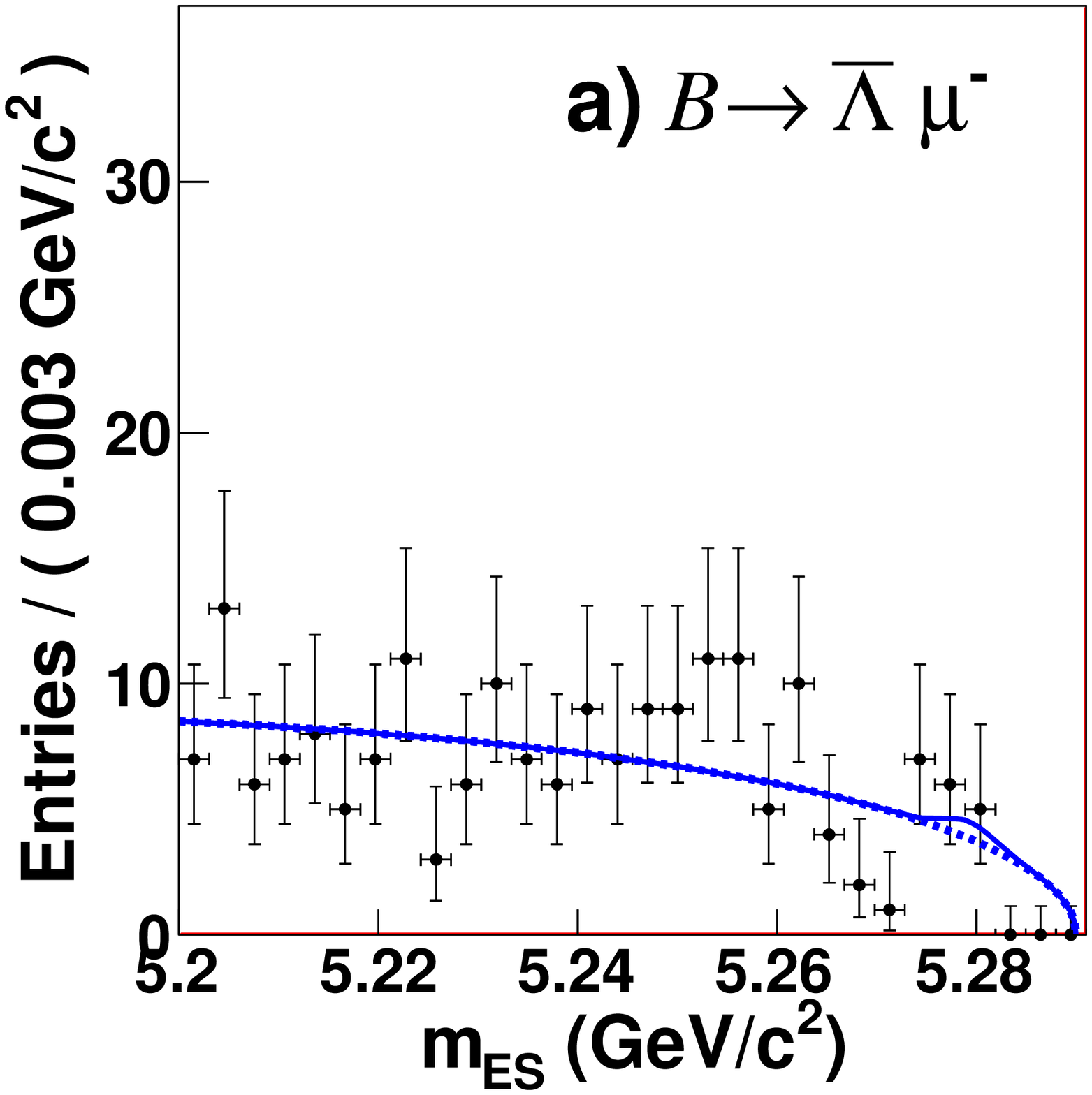}\hfill
\includegraphics[width=0.23\textwidth]{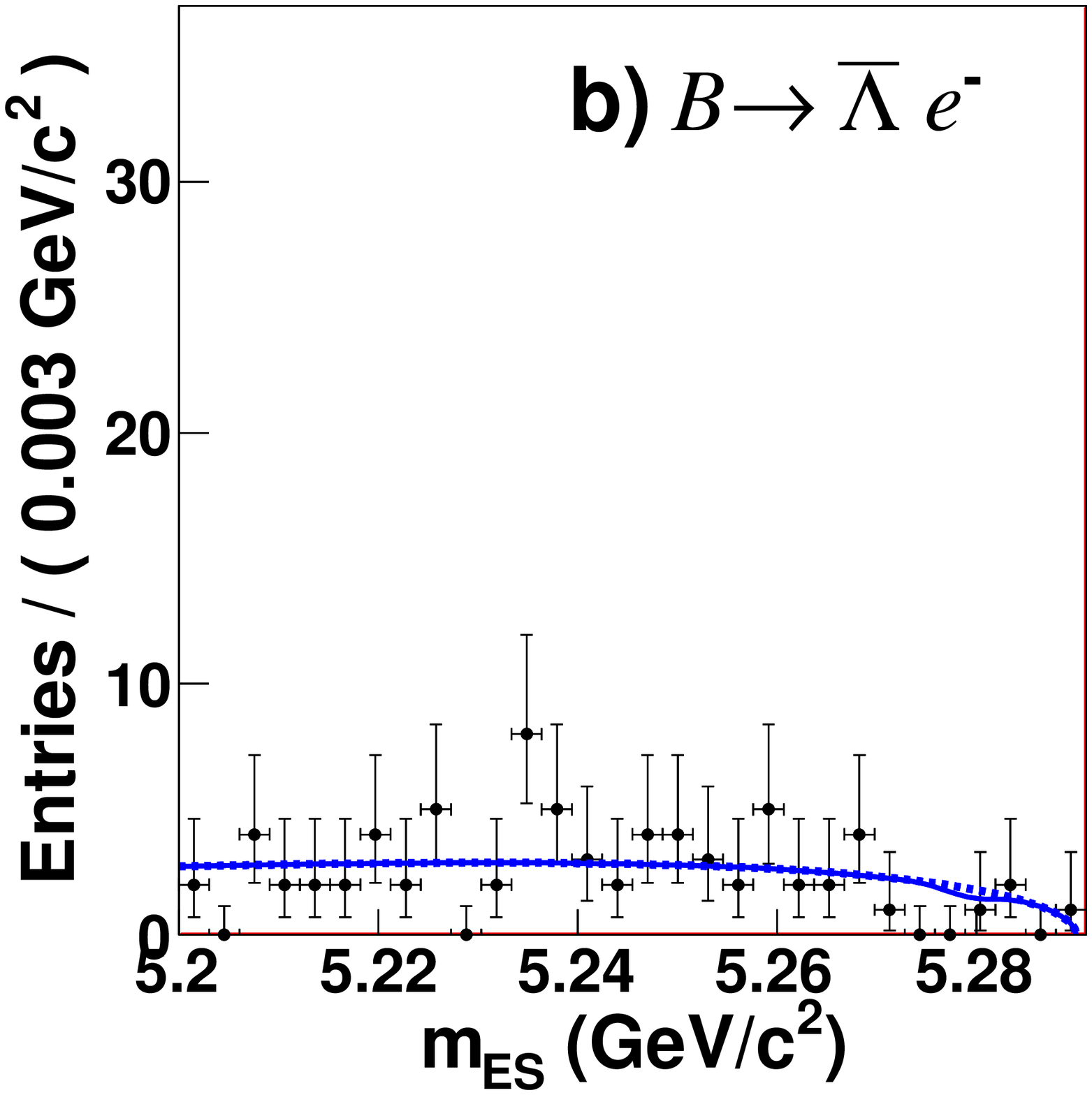}
\includegraphics[width=0.23\textwidth]{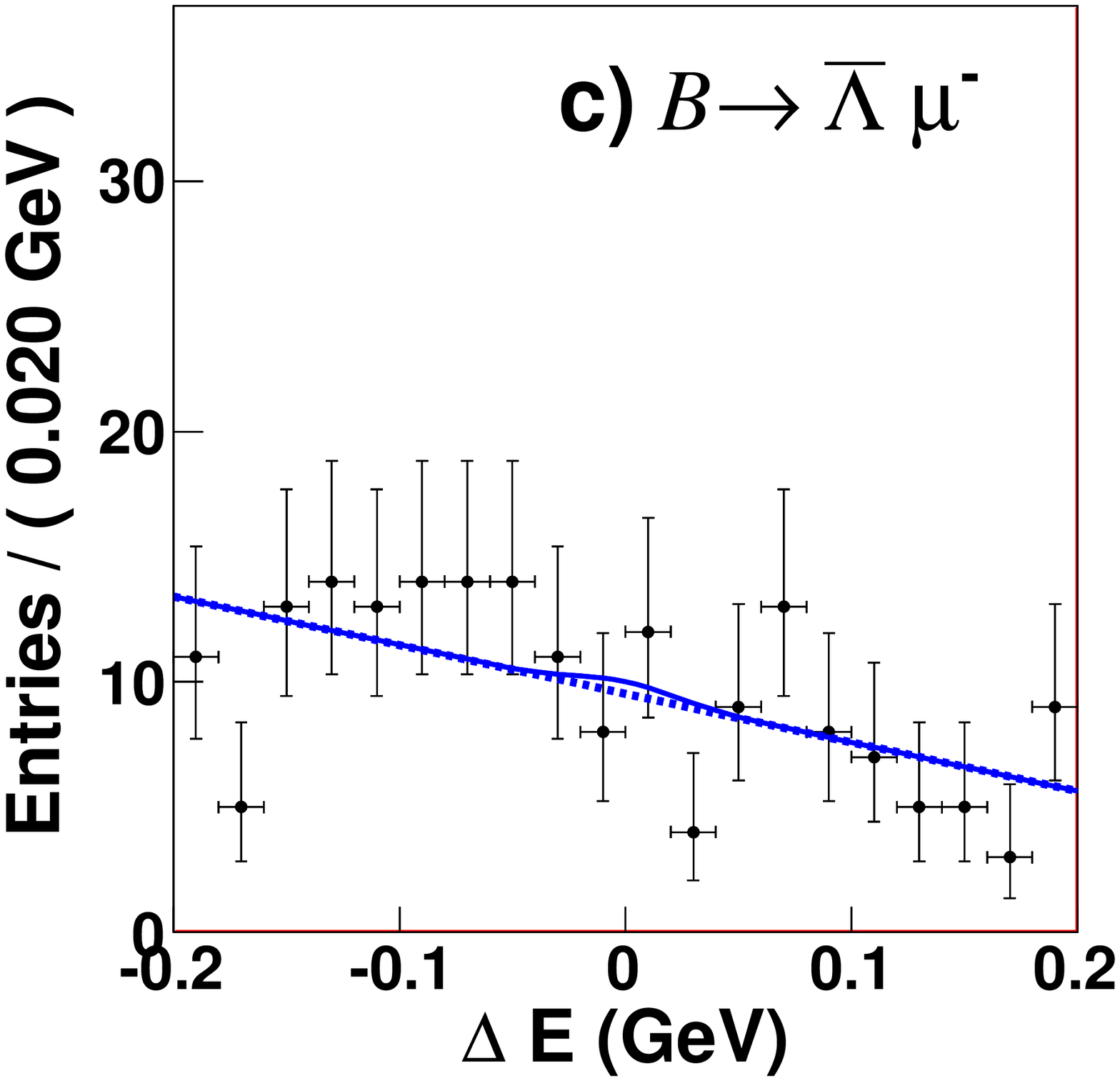}\hfill
\includegraphics[width=0.23\textwidth]{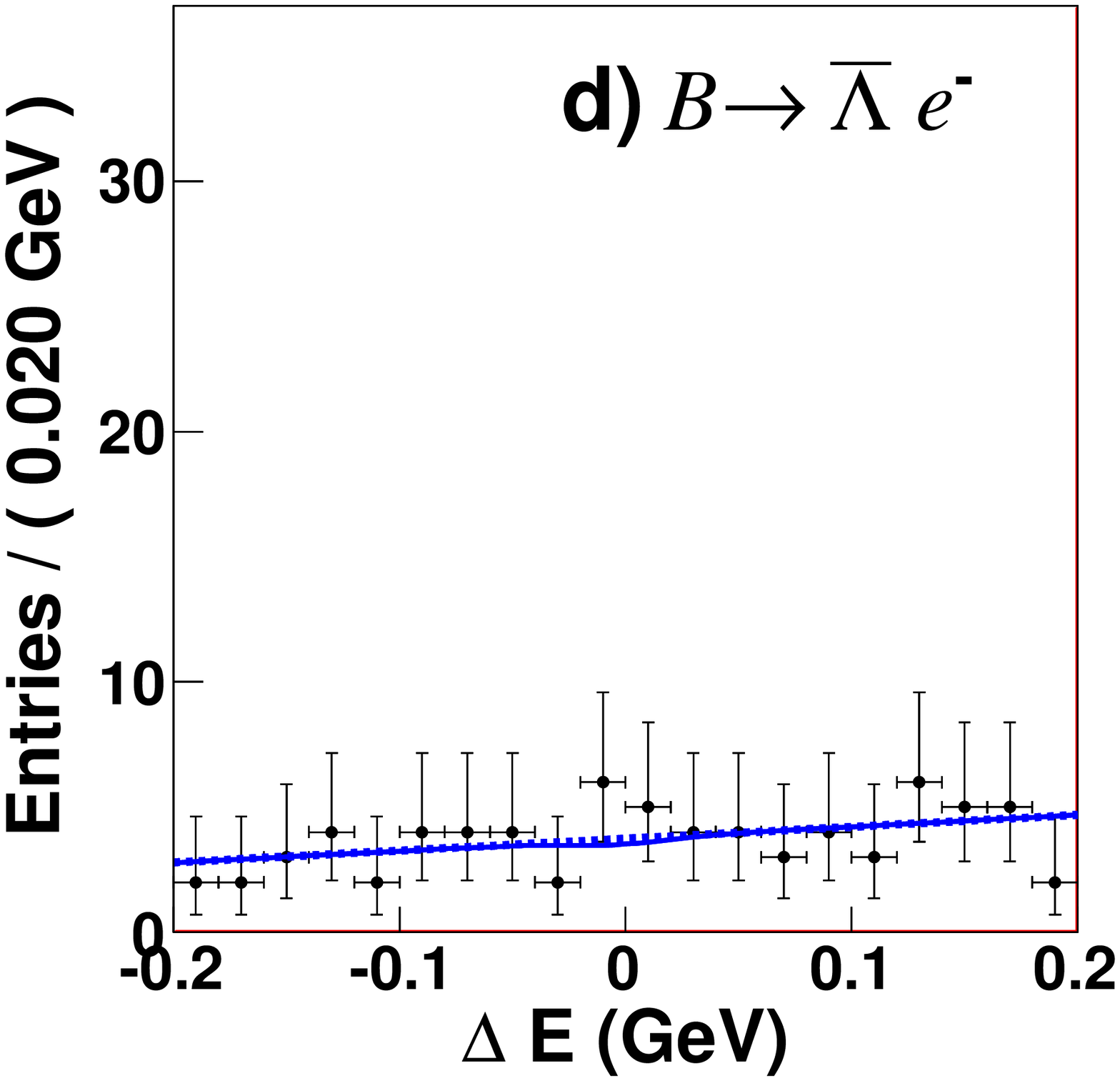}
\caption{Data for all events in the fitting region with overlaid fit results for $B^-\rightarrow\bar{\Lambda} \ell^-$ candidates. The left column
    is for the muon mode and the right column the electron mode.
    Distributions of (a,b) \mes and (c,d) \DeltaE are shown.
    Dashed lines represent the background components of the fit and solid
    lines represent the sum of the signal and background components. The lack of a significant signal
    makes the solid and dashed lines indistinguishable in some plots.\label{fig:data_al0}}
\end{center}
\end{figure}

The data and the fit projections are shown in 
Figs.~\ref{fig:data_lc},~\ref{fig:data_l0}, and~\ref{fig:data_al0} for
$B^0\rightarrow\Lambda_c^+\ell^-$, $B^-\rightarrow\Lambda\ell^-$,
and $B^-\rightarrow\bar{\Lambda}\ell^-$, respectively.

\section{\boldmath Results\label{sec:results}}

No significant signal is observed and an upper limit is calculated for
the branching fraction for each decay mode.  To calculate the upper
limit, the branching fraction is varied around the best fit value
$\mathcal{B}_{\rm best}$ and the other parameters are refit to map out
the difference in the $\ln$-likelihood: $\Delta \ln \mathcal{L} =  \ln
\mathcal{L}(\mathcal{B}_{\rm best})-\ln \mathcal{L}(\mathcal{B})$.  We
integrate the function $y = e^{-\Delta \ln \mathcal{L}}$ over $\mathcal{B}$.  While
the fit allows the branching fraction to assume negative values, we
ignore the unphysical region with $\mathcal{B}<0$ and calculate the
integral for $\mathcal{B}>0$.  We determine the value of the branching
fraction $\mathcal{B}_{90\%}$ for which
90\% of the area lies between $\mathcal{B}=0$ and $\mathcal{B}_{90\%}$
and interpret this as the upper limit at 90\% confidence level.
The results from the fit are given in Table~\ref{table:ul}.
Since the biases observed in Monte Carlo studies are small
compared to the statistical uncertainties,
we do not include their effect in the results.
For the $\bar{\Lambda} e^-$ decay mode, there are no candidates in the signal region. 
The fitted branching fraction for this decay mode is equal to the limit in the fit 
determined by the requirement that the PDF be positive throughout the fitting region.

\begin{table}[]
\caption{Total number of candidates used in the fit ($N_{\rm cand}$), 
the central value for the branching fraction returned by the fit ($\mathcal{B}$),
signal efficiency ($\epsilon$) excluding the contribution from the $\Lambda_{(c)}$ branching fraction, 
and upper limits on the branching fraction at 90\% confidence level ($\mathcal{B}_{90\%}$)
for each decay mode. 
\label{table:ul}}
\begin{center}
\begin{tabular}{l p{1cm} p{1.5cm} c c }
\hline
\hline
Decay mode & $N_{\rm cand}$ & $\mathcal{B}$ ($\times 10^{-8}$) & $\epsilon$ (\%) & $\mathcal{B}_{90\%}$ ($\times 10^{-8}$) \\
\hline
\rule[-0.1cm]{0.0cm}{0.6cm}$B^0 \rightarrow \Lambda_c^+ \mu^-$ & 814 & $-4_{-56}^{+71}$     & $26.3\pm0.9$ & $180$ \\
\rule[-0.1cm]{0.0cm}{0.6cm}$B^0 \rightarrow \Lambda_c^+ e^-$   & 651 & $190_{-90}^{+130}$ & $25.7\pm0.7$ & $520$ \\
\rule[-0.1cm]{0.0cm}{0.6cm}$B^-\rightarrow \Lambda \mu^-$       & 320 & $-2.3_{-2.5}^{+3.5}$ & $28.7\pm0.9$ & $6.2$ \\
\rule[-0.1cm]{0.0cm}{0.6cm}$B^-\rightarrow \Lambda e^-$         & 194 & $1.2_{-2.6}^{+3.7}$ & $27.2\pm0.6$ & $8.1$ \\
\rule[-0.1cm]{0.0cm}{0.6cm}$B^-\rightarrow \bar{\Lambda} \mu^-$ & 192 & $1.5_{-1.7}^{+2.6}$ & $31.3\pm1.0$ & 6.1 \\
\rule[-0.1cm]{0.0cm}{0.6cm}$B^-\rightarrow \bar{\Lambda} e^-$   & 74  & $-0.9_{-0.0}^{+0.7}$ & $30.0\pm0.6$ & $3.2$ \\
\hline
\hline
\end{tabular}
\end{center}
\end{table}


\section{Summary\label{sec:summary}}
Searches are performed for the decays $B^0\rightarrow\Lambda_c^+\ell^-$, 
$B^-\rightarrow\Lambda\ell^-$ and $B^-\rightarrow\bar{\Lambda}\ell^-$,
using the full \babar\ data set. 
No significant signal for any of the decay modes is observed
and upper limits are determined at the 90\% confidence level.


We are grateful for the excellent luminosity and machine conditions
provided by our \pep2\ colleagues, 
and for the substantial dedicated effort from
the computing organizations that support \babar.
The collaborating institutions wish to thank 
SLAC for its support and kind hospitality. 
This work is supported by
DOE
and NSF (USA),
NSERC (Canada),
CEA and
CNRS-IN2P3
(France),
BMBF and DFG
(Germany),
INFN (Italy),
FOM (The Netherlands),
NFR (Norway),
MES (Russia),
MICIIN (Spain),
STFC (United Kingdom). 
Individuals have received support from the
Marie Curie EIF (European Union),
the A.~P.~Sloan Foundation (USA)
and the Binational Science Foundation (USA-Israel).

\vfill
\bibliographystyle{ieeetr}

\begin{thebibliography}{99}
\bibitem{chg_conj}
Throughout this paper, whenever a decay mode is given, the charge conjugate is also
implied.
\bibitem{Lmoments}
$L$-moments are defined as $L_i\equiv \sum_j p_j^* \cdot | \cos \theta_j^* |^i/\sum_j p_j^*$
where $p_j^*$ and $\theta_j^*$ are the momentum and angle with respect to 
a given axis, respectively, for each particle $j$.
\end{thebibliography}

\end{document}